\magnification=1200
\baselineskip=18pt
\def\hs{\hskip}

\font\ninebx=cmbx9
\font\ninerm=cmr9
\font\ninemi=cmmi9

\font\ninesy=cmsy9
\font\eightrm=cmr8

\def\be{\begin{equation}}
\def\ee{\end{equation}} 
\def\ea{\end{array}\end{equation}}  
\def\bac{\begin{equation}\begin{array}{rll}}
\def\uq{U_q (\widehat{sl(2)})}

\def\ep{\epsilon}
\def\ra{\rightarrow}


\line{December, 1996\hfil Preprint ITP-SB-96-72}
\bigskip\bigskip
\centerline{\bf EXACT FOUR-SPINON DYNAMICAL CORRELATION} 
\centerline{\bf FUNCTION OF THE HEISENBERG MODEL}
\bigskip
\centerline{A. Abada$^{1}$, A.H. Bougourzi$^{2}$ and B. 
Si-Lakhal$^{1}$}
\bigskip
\centerline{$^1$ \it D\'epartement de Physique, 
Ecole Normale Sup\'erieure,}
\centerline{\it BP 92, Vieux-Kouba, 16050 Alger, Algeria}
\medskip
\centerline{$^{2}$\it Institute for Theoretical Physics}
\centerline{\it SUNY at Stony Brook, Stony Brook, NY 11794}

\bigskip\bigskip\bigskip\bigskip

\centerline{\ninebx Abstract}
\bigskip
\hbox{\hs1truein\vbox{\hsize=4.5truein\baselineskip=12pt\ninerm 
In this paper we derive the exact expression of the four-spinon
contribution to the dynamical correlation function of the
spin $\ninesy\ninemi S= 1/2$ anisotropic ($\ninemi XXZ$) 
Heisenberg model
in the antiferromagnetic regime. We extensively study its 
isotropic
($\ninemi XXX$) limit and derive perturbatively the Ising one. 
Our method relies on the quantum affine symmetry of the model, 
which allows
for a systematic diagonalization of the Hamiltonian in the 
thermodynamic
limit and for an exact calcualtion of matrix elements 
of local spin
operators. In fact, we argue that the familiar criticism 
of this method 
related to the complication of these matrix elements 
is not justified.
First, we give, in the form of contour integrals, an exact
expression for the n-spinon contribution. After we 
compile recently
found results concerning the two-spinon contribution,
we specialize the n-spinon formula to the new case 
$\ninemi\ninesy n=4$.
Then
we give an explicit series representation of this 
contribution in the
isotropic limit. Finally, after we show that this 
representation is free
of divergences, we discuss the Ising limit in which a simple
expression is found up to first order in the 
anisotropy parameter.}}

\vfil\break

\line{\bf  Introduction\hfil}
\bigskip
Because of their well pronounced strong 
quantum behavior and highly
nontrivial many-body interactions investigated in neutron
scattering experiments on ferromagnetic 
compounds, and because of their
rich mathematical structures, quantum spin 
chains have been the subject
of intensive study over the past seven 
decades [1-19]. In this respect, 
one important quantity that plays a crucial role in these
magnetic neutron scattering experiments is 
the dynamical correlation
function (DCF) of local spin operators. 
This is because the cross
section of such a scattering process is directly proportional
to this DCF. In the past, most of the 
theoretical work related to such
experiments has relied on approximation 
schemes [20-22].

However, certain one-dimensional 
compounds can be adequately described
by models like the anisotropic Heisenberg ($XXZ$) chain,
which has been shown to be amenable to exact manipulations 
[7,14]. Indeed, many static physical 
quantities of the $XXZ$ chain
have been determined exactly. But in contrast, dynamical
physical quantities, and in particular 
the DCF, have remained elusive, due
mainly to the complicated structure 
of the Hilbert space, which is
built on "spinons" with nontrivial statistics and interactions. 
This is particularly the case in the antiferromagnetic regime and
the thermodynamic limit. Now as we understand it, the main reason
why such a model is tractable exactly is its invariance under the
quantum affine symmetry. The recognition 
and systematic exploitation
of this symmetry has led to the exact diagonalization in the 
thermodynamic
limit of the corresponding Hamiltonian [23]. 
The Hilbert space is well
defined and consists of multi-spinon 
particles. Moreover, using the
technique of bosonization of the 
corresponding quantum affine algebra 
[24,25,26], one is able to calculate exactly 
matrix elements of local spin
operators. This is explicitly carried out by Jimbo and Miwa in 
[27].

Nevertheless, one main criticism plaguing such 
calculations is that the
matrix elements of local operators are typically 
expressed in the form of
complicated contour integrals which are hardly useful 
for practical
numerical applications related to experimental 
data and/or exact finite-chain
calculations. In fact, this is 
precisely the reason why most of the very
recent developments have been restricted to the derivation, 
using these matrix elements, of only the two-spinon contribution
to the transverse DCF of the $XXZ$ chain and its 
$XXX$ and Ising limits:
the complicated contour integrals are simply 
not present [28-30].
Note however that in [31],
the two-spinon longitudinal DCF, where two such 
contour integrals arise,
has been investigated perturbatively around the Ising limit.
As shown in Ref. [29], in the $XXX$ limit, the 
two-spinon contribution accounts for much of but {\it not all} 
the total DCF. 
Therefore, it is natural to try to look into the next 
contribution,
namely that of four spinons. As far as we know, no numerical 
or analytic
result, even approximate, related to this latter is yet 
available in the
literature. The main reason is that in any approximation
scheme, it is exremely difficult to single out from the rest the
contribution of just the four spinons, and this even for a 
finite chain.
In fact, to the best of our knowledge, the first attempt to 
tackle in
a systematic way the case $n>2$ is ref. [32], where a 
formal expression
for the exact n-spinon contribution to the DCF of the 
$XXX$ chain is
presented. But the contour integrals that arise there are 
not investigated
thoroughly.

It is the purpose of this paper to determine the exact 
expression of
the four-spinon contribution to the DCF of the $XXZ$ model in the
antiferromagnetic regime and in the thermodynamic limit, 
and to discuss its isotropic $XXX$ and Ising
limits. In this contribution, a single contour integral 
arises with an
infinite number of simple poles, integral that we express in 
the form of
an infinite series. Even thus expressed, the four-spinon
contribution in the $XXZ$ model is still quite complicated, 
whereas
its isotropic and Ising limits simplify considerably. 
Because of this, we concentrate  our study on the $XXX$ and 
Ising cases.
In the $XXX$ case, we show that our expression is free of 
potential
infinities, and thus is well defined. In the Ising case,
we derive a simple expression for the four-spinon 
contribution up to
first order in the anisotropy parameter. We then compare it 
to the two-spinon contribution and discuss some 
discrepancies with results
found in the literature.

The paper is organized as follows. In the next section, 
following [23,27],
we review the diagonalization in the 
thermodynamic limit of the $XXZ$
model in the antiferromagnetic regime and, in 
particular, briefly explain
how it allows for an exact calcualtion of the 
matrix elements of local
spin operators. In section 2, we define the DCF and use the
completeness relation to write it as a sum 
over n-spinon contributions,
$n$ even, for which we derive a general 
coutour representation. In section 3,
For the sake of completeness, we specialize to 
the case $n=2$ and rederive
all known results for the two-spinon 
contribution and update some of them.
In section 4, we carry out a systematic study
of the case $n=4$ in the isotropic limit and 
discuss the Ising one.
Finally, we devote section 5 to our conclusions.
\bigskip\bigskip

\line{\bf 1. The anisotropic Heisenberg model in 
the antiferromagnetic
regime\hfil}
\bigskip
In this section, we briefly review the diagonalization in the
thermodynamic limit of the anisotropic Heisenberg $XXZ$ model 
in the
antiferromagnetic regime. We follow Refs. [23,27] where the 
quantum affine
symmetry present in the model is exploited. This model is 
defined by
the Hamiltonian:
$$
 {H_{XXZ}}=
 -{1\over 2} \sum_{n=-\infty}^{\infty} (\sigma_n^x 
\sigma_{n+1}^x 
 +\sigma_n^y \sigma_{n+1}^y + \Delta\sigma_n^z \sigma_{n+1}^z)\ ,
 \eqno(1.1)
$$ 
where $\Delta=(q+q^{-1})/2$ is the anisotropy parameter. 
We consider the model in the antiferromagnetic
regime characterized by $\Delta <-1$ or equivalently by $-1<q<0$.
Here $\sigma_n^{x,y,z}$ 
are the usual
Pauli matrices acting non-trivially at the 
$n^{\rm th}$ position of the 
formal 
infinite tensor product 
$$
W= \cdots V \otimes V \otimes V  \cdots\ ,\eqno(1.2) 
$$ 
where $V$ is the two-dimensional representation of $U_q(sl(2))$ 
quantum group, with basis $\{ v_+, v_-\}$. 
The  main point of Refs. [23,27]  
is that the action of $H_{XXZ}$ 
on $W$ is
not well defined due to the appearance of 
divergences. 
However, because this model is symmetric 
under the quantum group $\uq$, its 
 Hilbert space, which is free of divergences,
is identified with the following level 0 $\uq$ 
module:
$$
{\cal F}=\sum_{i,j}V(\Lambda_i)\otimes V(\Lambda_j)^*.\eqno(1.3) 
$$
Here $\Lambda_i$ and $V(\Lambda_i); i=0,1$ are  level 
1 $\uq$-highest weights and $\uq$-highest weight modules, 
respectively, whereas
$V(\Lambda_i)^*$ are  dual modules defined from  
$V(\Lambda_i)$  
through the antipode.
The module $V(\Lambda_i)$ is identified 
with the
subspace of the following formal semi-infinite tensor 
product of $V$'s:
$$
 X= \cdots V \otimes V \otimes V,\eqno(1.4)
$$ 
and which consists  of all linear combinations of  
spin configurations with  fixed boundary conditions such that 
the eigenvalues of 
$\sigma^z_n$ are $(-1)^{i+n}$ in the limit $
 n\ra -\infty$. Similarly, the module $V(\Lambda_i)^*$ is 
associated with the right semi-infinite tensor product of the
$V$'s. What is impressive about this method is that it allows
for a Fock space (or spinon-particule  picture) 
representation of the Hilbert space, that is, this
space can be created from a vacuum state through the
successive actions of creation operators. 
This is achieved in terms
of what is called type II vertex operators $\Psi^{1-i,i}(\xi)$
and which interwine the $\uq$ modules as:
$$
\Psi^{1-i,i}(\xi):\qquad V(\Lambda_i)\ra 
V(\xi)\otimes V(\Lambda_{1-i}),\eqno(1.5)
$$    
where $V(\xi)$ is the affinization of the two-dimensional 
representation
$V$ in such a way that it is isomorphic to 
$V\otimes C[\xi,\xi^{-1}]$, with $C[\xi,\xi^{-1}]$ being the
algebra of polynomials in $\xi$ and $\xi^{-1}$. Here, $\xi$ is a
spectral parameter useful in parameterizing the energies
and momenta of the spinons, which are by definition the
eigenstates of the Hamiltonian. This spectral parameter lies 
on the unit circle. More specifically, the
creation and annihilation operators are constructed as follows:
decompose $\Psi^{1-i,i}(\xi)$ as
$$
\Psi^{1-i,i}(\xi)=\sum_{\epsilon=\pm 1}v_\epsilon \otimes 
\Psi_\epsilon ^{1-i,i}(\xi),\eqno(1.6)
$$ 
and let $(\Psi^{1-i,i}(\xi))^*$ be the conjugate operator to
$\Psi^{1-i,i}(\xi)$. Then the following important 
commutation  relations have been shown in Ref. [23]:
$$
\eqalign{
[H_{XXZ},\Psi^{1-i,i}(\xi)]&=-e(\xi) \Psi^{1-i,i}(\xi)\ ;\cr
[H_{XXZ},(\Psi^{1-i,i}(\xi))^*]&=e(\xi)
(\Psi^{1-i,i}(\xi))^*\ ,\cr}\eqno(1.7)
$$
and which are the usual definitions of annihilation and 
creation operators respectively.  Moreover, the vacuum state
in the sector $i$ is defined to be
$$
|0>_i=c_1^{-1/2}(-q)^{i/2-\rho}P_i,\eqno(1.8)
$$
where $P_i$ is the projector of $V(\Lambda_0)\oplus V(\Lambda_1)$
on $V(\Lambda_i)$, 
$\rho$ is the grading element of $\uq$, 
and $c_1=(q^2;q^4)_\infty$ (see relation
(1.13) for the definition of $(y;x)_\infty$). 
The multi-spinon eigenstates are then systematically 
constructed as
$$
|\xi_n,\cdots 
\xi_1>_{\ep_1,\cdots \ep_n;i}=
c_2^{-n/2}\Psi^{*}_{\epsilon_n}(\xi_n) \cdots 
\Psi^{*}_{\epsilon_1}(\xi_1)|0>_i.\eqno(1.9)
$$
Here $c_2=(q^2;q^4)_\infty/(q^4;q^4)_\infty$ and for 
clarity we have omitted the indices refering to the
sectors on which the $\Psi^{*}_{\epsilon_j}(\xi_j)$'s  are
acting, and the $\epsilon$'s are the projections of the
$z$-component of the total spin. 
The completeness relation in ${\cal F}$ reads then [27]
$$
 {\bf I}=\sum_{i=0,1}\sum_{n \geq 0} \sum_{\ep_1,
\cdots,\ep_n=\pm 1}
 {1 \over {n !}} \oint  {d\xi_1\over 2\pi i \xi_1} 
\cdots {d\xi_n\over
 2\pi i \xi_n} |\xi_n,\cdots,\xi_1>_{{\ep_n,\cdots,\ep_1};\; i}\;
 {_{i;{\ep_1,\cdots,\ep_n}}{<\xi_1,\cdots,\xi_n|}}. \eqno(1.10)
$$
The actions of $H_{XXZ}$ and the translation operator $T$ 
(which shifts the
spin chain by one site) on the spinon states are given by:
$$
 \eqalign{
 T|\xi_1,\cdots,\xi_n>_i &=\prod_{i=1}^n\tau(\xi_i)^{-1}
 |\xi_1,\cdots,\xi_n>_{1-i}\ ,\quad 
 T|0>_i =|0>_{1-i}\ ; \cr
 H_{XXZ}|\xi_1,\cdots,\xi_n>_i 
 &=\sum_{i=1}^n e(\xi_i)|\xi_1,\cdots,\xi_n>_i\ ,\cr} \eqno(1.11)
$$  
where:
$$
 \eqalign{ \tau(\xi)&= \xi^{-1} 
 {\theta_{q^4} (q \xi^2) \over \theta_{q^4} (q \xi^{-2})}=
 e^{-i p(\alpha)}\ ,
 \quad p(\alpha)=am({2K\over \pi}\alpha)+{\pi/2}\ ; \cr
 e(\xi) &={ 1-q^2 \over 2 q} \xi {d \over d \xi}\log \tau(\xi)=
 {(q-q^{-1})K\over\pi}\sqrt{1-k^2\cos^2(p)}\ .\cr} \eqno(1.12)
$$
In the above equations, $e$ and  $p$ are the energy and the 
momentum of the spinon respectively,  $am(x)$ is the usual 
elliptic amplitude function with the complete elliptic integrals
$K$ and $K^\prime$, and moduli $k$ and $k^\prime$. 
Also:
$$
 \eqalign{q&=-\exp(-\pi K^\prime/K)\ ;\cr
 \xi&=ie^{i\alpha}\ ;\cr
 \theta_x(y)&=(x;x)_{\infty} (y;x)_{\infty} 
 (x y^{-1};x)_{\infty}\ ;\cr
 (y;x)_{\infty}&=\prod_{n=0}^{\infty} (1-y x^n)\ .\cr}
\eqno(1.13)
$$
Furthermore, $\sigma^{x,y,z}_n(t)$ at time $t$ and 
position $n$ is related
to $\sigma^{x,y,z}_0(0)$ at time $t=0$ and position $n=0$ via:
$$
 \sigma^{x,y,z}_n(t)=\exp(i tH_{XXZ} ) T^{-n}
 \sigma^{x,y,z}_0(0) T^{n}\exp(-i tH_{XXZ} )\ . \eqno(1.14)
$$
At this stage, one should have expected the  
existence of 
 ``type I" vertex operators. Indeed, the latter denoted by
$\Phi^{1-i,i}(\xi)$ are also interwiners of the same modules
as those of type II vertex operators but in opposite order. 
Let us remind
that the order of modules appearing in tensor produtcs is
relevant in non-cocommutative algebras such as $\uq$. More 
precisely, they intertwine these $\uq$ modules as:
$$
\Phi^{1-i,i}(\xi):\qquad V(\Lambda_i)\ra 
V(\Lambda_{1-i})\otimes V(\xi).\eqno(1.15)
$$
They are similarly decomposed as
$$
\Phi^{1-i,i}(\xi)=\sum_{\epsilon=\pm 1} 
\Phi_\epsilon^{1-i,i}(\xi)\otimes v_\epsilon.\eqno(1.16)
$$ 
``La raison d'\^etre" of such type I vertex operators is that
they allow for the translation of the action of 
 local operators from that on
the space $W$ to  one on ${\cal F}$. To be specific, any local
operator can be written in terms of two-dimensional unit
matrices $E_{\mu_1\mu_2}$, which in turn can be represented
in ${\cal F}$ as:
$$
E_{\mu_1\mu_2}=c_2\Phi^*_{\mu_1}(1)\Phi_{\mu_2}(1)\otimes {\rm 
id}.\eqno(1.17)
$$
Using the properties of the scalar products defined 
on $\uq$-highest weight modules  it has been shown in
Ref. [23] that the matrix elements  of local operators can
be then expressed  as traces over highest weight 
modules of products of type I and type
II vertex operators. For example the matrix element of the
unit local operator $E_{\mu_1\mu_2}$ acting on site 1 is
equal to
$$
_{i}<0|E_{\mu_1\mu_2}|\xi_n,\cdots,\xi_1>_{\epsilon_n,\cdots,
\epsilon_1; i}=c_2^{1-n/2}{{\rm Tr}_{V(\Lambda_i}
\bigl(q^{-2\rho+i}\Phi^*_{\mu_1}(1)\Phi^*_{\mu_2}(1)
\Psi^*_{\epsilon_n}(\xi_n)\cdots \Psi^*_{\epsilon_1}(\xi_1)\bigr)
\over {\rm Tr}_{V(\Lambda_i}
\bigl(q^{-2\rho+i}\bigr)}.\eqno(1.18)
$$
Unfortunately, these traces are over infinite-dimensional modules
and hence cannot be easily evaluated. What is so
special about the anisotropic Heisenberg model in the
antiferromagnetic regime is that the quantum affine algebra
$\uq$, its highest weight modules $V(\Lambda_i)$,
 and both its type I and type II vertex operators can be
bosonized in terms of a set of simple harmonic oscillators 
satisfying 
the usual Heisenberg algebra\footnote{*}{Heisenberg not 
only defined the model, he also defined the right tool for
its exact resolution, though he did not realize it 
at that time}. 
Therefore, this problem is translated to a familiar 
quantum field theory
problem of computing the Green functions of a free 
bosonic field. Using the Wick theorem and
other procedures the above traces have been explicitly computed
by Jimbo and Miwa [27]. From this brief summary it is tempting
to postulate that whether a model is explicitly integrable 
or not depends on whether it is bosonizable or not.

\bigskip\bigskip

\line{\bf 2. The n-spinon transverse DCF of the $XXZ$ model\hfil}
\bigskip
First, we define one of the two (equal) transverse components
of the DCF in the case of the  $XXZ$ model. 
The reason why we focus only on the transverse DCF is 
that it involves
the least number of contour integrals and, in particular, 
the transverse
two-spinon one does not involve any. This is to be 
contrasted with
the longitudinal DCF which, even in the simplest case 
of two spinons,
involves two complicated contour integrals and this number 
increases
with  the number of spinons. We define this component as:
$$
 S^{i,+-}(w, k)=\int_{-\infty}^{\infty} dt \sum_{m\in Z}
 e^{i(wt+km)} {_i}< 0|\sigma^+_m(t)\sigma^-_0(0)|0>_i\ , 
\eqno(2.1)
$$
where $w$ and $k$ are the neutron energy and momentum transfer
respectively, and $i$ corresponds to the boundary condition. 
Using the completeness relation, the n-spinon contribution 
is given by:
$$
 \eqalign{&S_n^{i,+-}(w,k)= {2\pi\over n!} \sum_{m\in Z} 
\sum_{\ep_1,\dots,
 \ep_n} \oint {d\xi_1\over 2\pi i \xi_1}
\dots{d\xi_n\over 2\pi i \xi_n}
 e^{im\big(k+\sum_{j=1}^n p(\xi_j)\big)}
 \delta\big(w-\sum_{j=1}^n e(\xi_j)\big)\cr
 &\qquad\times {_{i+n}<0|}\sigma^+_0 
 (0)|\xi_n,\dots ,\xi_1>_{\ep_n,\dots,\ep_1;\; i+n}\>
 {_{i;\ep_1,\dots ,\ep_n}<\xi_1,\dots , \xi_n|}
 \sigma^-_0(0)|0>_i\ .\cr}\eqno(2.2)
$$
It can be rewritten in the following more tractable way:
$$
 \eqalign{&S_n^{i,+-}(w,k)={2\pi\over n!}\sum_{\ep_1,\dots ,
\ep_n} 
 \oint {d \xi_1\over 2\pi i \xi_1}\dots { d \xi_n\over 2 \pi i 
\xi_n} 
 \sum_{m\in Z} e^{2mi\big(k+\sum_{j=1}^n p(\xi_j)\big)}
 \delta\big(w-\sum_{j=1}^n e(\xi_j)\big)\cr
 &\qquad\times \Big({_{i}<0|}\sigma^+_0(0)|\xi_n,\dots ,
 \xi_1>_{\ep_n,\dots ,\ep_1;\; i}\quad{_{i;\ep_1,\dots ,\ep_n}
 <\xi_1,\dots ,\xi_n|}\sigma^-_0{(0)}|0>_i \cr
 &\qquad\qquad +e^{i\big(k+\sum_{j=1}^n p(\xi_j)\big)}\;
 {_{1-i}<0|}\sigma^+_0(0)|\xi_n,\dots ,\xi_1>_{\ep_n,\dots ,
\ep_1;1-i}\cr
 &\qquad\qquad\qquad\qquad\qquad {_{i;\ep_1,\dots ,\ep_n}
 <\xi_1,\dots ,\xi_n|}\sigma^-_0{(0)}|0>_i \Big)\ .\cr}\eqno(2.3)
$$
The non-vanishing matrix elements have been computed in [27].
In particular, they satisfy the following relations:
$$
 \eqalign{{_i{<}}0|\sigma^-_0(0)|\xi_n,\dots,\xi_1>_{\ep_n,
\dots,\ep_1;\; i}
 &={_{1-i}{<0|}}\sigma^+_0(0)|\xi_n,\dots,\xi_1>_{-\ep_n,\dots,
 -\ep_1;\; 1-i}\cr
 &={_{i}{<0|}}\sigma^+_0(0)|-q\xi_1^{-1},\dots,-q\xi_n^{-1}>_{
 -\ep_1,\dots,-\ep_n;\; i}\ ;\cr
 {_{i;\ep_1,\dots,\ep_n}{<}} \xi_1,\dots,
\xi_n|\sigma^-_0(0)|0>_i&=
 {_i{<0|}}\sigma^-_0(0)|-q \xi_1,\dots,-q \xi_n>_{-\ep_1,
 \dots,\ep_n;\; i}\ .\cr}\eqno(2.4)
$$
We ease slightly the notation by writing:
$$
 X^i_{\ep_n,\dots,\ep_1}(\xi_n,\dots ,\xi_1) \equiv
 {_{i}{<0|}}\sigma^+_0(0)|\xi_n,\dots ,\xi_1>_{\ep_n,\dots, 
\ep_1;\; i}\ ,
 \eqno(2.5)
$$
and we have:
$$
 \eqalign{X^i_{\ep_n,\dots,\ep_1}(\xi_n,\dots ,\xi_1)&=
 \bigl(\delta_{\sum_{j=1}^n \ep_j,-2}\bigr)
 {\prod_{1\leq j< j^\prime\leq n} \gamma(u_{j^\prime}/u_j)\over
 \prod_{j=1}^n (-qu_j;q^4)_\infty (-q^3 u^{-1}_j;q^4)_\infty}\cr
 & \prod_{j=1}^n \xi_j^{j-(1+\ep_j)/2-i}
 \prod_{\ell\in\L} \oint_{C_\ell} 
 {dv_\ell\over 2 \pi i v_\ell}
 \prod_{j<\ell }\big(q v_\ell^{-1}-q^{-1} u_j^{-1})\big)
 \prod_{j>\ell}\big(v_\ell^{-1}- u_j^{-1})\big)\cr
 & \prod_{j,\ell}{1\over (u_j/v_\ell;q^4)_\infty
 (q^{-2} v_\ell/u_j;q^4)_\infty }
 \prod_{\ell} (-q^{-1}v_\ell;q^2)_\infty 
(-q^3/v_\ell;q^2)_\infty \cr
 &\prod_{\ell<\ell^\prime}v_{\ell^\prime}
 (v_{\ell^\prime}-q^{-2}v_\ell) (v_\ell/v_{\ell^\prime};
 q^2)_\infty (q^2 v_{\ell^\prime}/v_{\ell};q^2)_\infty 
 G^+_i(v,u)\ .\cr}\eqno(2.6)
$$
In the above equation, we have used the following  definitions:
$$
 \eqalign{u_j&=-\xi^2_j\ ;\cr
 G^+_i(v,u)&=C^+_{n,i}\prod_{\ell\in\L}(-v_\ell)^i
 \theta_{q^8}\Big(-q^{2-n+4i}{\prod_{\ell} v_\ell^2 \over
 \prod_{j} u_j} \Big)\ ;\cr
 C^+_{n,i}&=\delta_{t,{n\over 2}-1} 
 (-q)^{-{n^2\over 4}+{(3-i)n\over 2}-1} (1-q^{-2})^{{n\over 2}-1}
 (q^2;q^4)_\infty (q^4;q^4)^{n-1}_{\infty}\rho^n(q)\ ;\cr
 \gamma(\xi)&={(q^4\xi;q^4;q^4)_{\infty}
 (\xi^{-1};q^4;q^4)_{\infty}\over  
 (q^6\xi;q^4;q^4)_{\infty}
 (q^2\xi^{-1};q^4;q^4)_{\infty}}\ ;\cr
 \rho(q)&={(q^4;q^4;q^4)_{\infty}\over
 (q^6;q^4;q^4)_{\infty}}\ .\cr} \eqno(2.7)
$$
Moreover, $t$ is the number of elements in the set $\L$, 
which is defined by:
$$
 \L=\{j,\quad {\rm s.t.} \quad \ep_j=+1,\quad {\rm and}\quad
 \sum_{i=1}^n \ep_i=-2,\quad 1\leq j\leq n\}\ .\eqno(2.8)
$$ 
Finally, in (2.7), each contour $C_\ell$ includes the poles
$q^{4n}u_j,\> n\geq 0$ only, excluding the poles 
$q^{2-4n}u_j,\> n\geq 0$.

Now, restricting ourselves to the first Brillouin zone (i.e., 
$0\leq k\leq 
2\pi$), integrating the delta functions, and keeping track 
of the Jacobian
factors, we find:
$$
 S^{i,+-}_n(w,k-\pi)=C_n \oint{d\xi_3\over 2\pi i \xi_3}\dots
 {d\xi_n\over 2\pi i \xi_n}\sum_{(\xi_1,\xi_2)}
 {h^i(\xi_n,\dots,\xi_3,\xi_2,\xi_1)\over
 J(p_1,p_2)}\ ,\eqno(2.9)
$$
where we have:
$$
 \eqalign{h^i(\xi_n,\dots,\xi_1)=&
 \sum_{\ep_n,\dots,\ep_1}\big(X^i_{\ep_n,\dots,\ep_1}(\xi_n,
\dots ,\xi_1)
 -X^{1-i}_{\ep_n,\dots,\ep_1}(\xi_n,\dots ,\xi_1)\big)\cr
 & \times X^i_{\ep_n,\dots,\ep_1}(\xi_n^{-1},\dots ,\xi_1^{-1})
\ ;\cr
 J(p_1,p_2)=&|\sin(2p_1)\sqrt{1-k^2\cos^2(p_2)}-
 \sin(2p_2)\sqrt{1-k^2\cos^2(p_1)}|\ ;\cr
 C_n  =& {\pi^3\over 4n! (q-q^{-1})k^2 K^3}\ .\cr}\eqno(2.10)
$$
In equation (2.9), the sum over the pairs $(\xi_1,\xi_2)$ is a 
sum
over all the solutions to the energy-momentum conservation laws:
$$
 \eqalign{w&=e(\xi_1)+e(\xi_2)+ \sum_{j=3}^n e(\xi_j)\ ;\cr
 k&=-p(\xi_1)-p(\xi_2)- \sum_{j=3}^n p(\xi_j)\ .\cr}\eqno(2.11)
$$
The physical DCF is obtained by summing over both
sectors $i=0,1$. 
Finally, let us mention that the isotropic limit of this integral
representation of the n-spinon contribution has been 
presented in Ref. [32].

\bigskip\bigskip

\line{\bf 3. The two-spinon DCF and its isotropic and 
Ising limits\hfil}
\bigskip
In this section, we discuss the case $n=2$. This is 
particularly interesting 
because no contour integral arises, that is, the set $\L$ 
in equation (2.8)
above is empty. We briefly rederive the known results 
[28-30] and,
more importantly, comment on some new properties of theirs.
The two-spinon DCF takes the following simple form:
$$
 S^{i,+-}_2(w, k-\pi)=
 C_2  {h^i(\xi_1,\xi_2)\over
 J\big(p_1,p_2\big)}\ .\eqno(3.1)
$$
Note that $S_2$ vanishes outside 
the domain of 2-spinon continuum set of
states.  The rest of the notations is as follows:
$$
 \eqalign{ h^i({\xi_1},{\xi_2})&=
 \big(X^i(\xi_1,\xi_2)-X^{1-i}(\xi_1,\xi_2)\big)\;
 X^{i}(\xi^{-1}_1,\xi^{-1}_2) \cr
 &={\gamma_{-}\big({\xi}_2^2/{\xi}_1^2\big)
 \gamma_{-}\big({\xi}_1^2/{\xi}_2^2\big)
 \gamma_{+}(q^{-2}) (q^2;q^2)^4
 Y^i(\xi_1,\xi_2)
 \over \theta_{q^2}(q\xi_1^{2}) \theta_{q^2}(q\xi_2^{2})}\ ;\cr
 Y^i(\xi_1,\xi_2)&=
 q^{2(1-i)}\theta_{q^8}(-\xi_1^{2}\xi_2^{2}q^{4i})
 (\theta_{q^8}(-\xi_1^{-2}\xi_2^{-2}q^{4i})
 +(q\xi_1\xi_2)^{2i-1}
 \theta_{q^8}(-\xi_1^{-2}\xi_2^{-2}q^{4(1-i)}))\ ;\cr
 \gamma_{\pm}(\xi)&={(\xi q^4(-q)^{1\pm 1};q^4;q^4)_{\infty}
 (\xi^{-1}(-q)^{1\pm 1};q^4;q^4)_{\infty}\over
 (\xi q^4(-q)^{3\pm 1};q^4;q^4)_{\infty}(\xi^{-1}(-q)^{3\pm 1};
 q^4;q^4)_{\infty}}\ ;\cr
 J(p_1,p_2)&=
 |\sin\big(2p_1\big)
 \sqrt{1-k^2\cos^2\big(p_2\big)}- \sin\big(2p_2\big)
 \sqrt{1-k^2\cos^2\big(p_1\big)}|\ ;\cr
 C_2&={\pi^3\over 2(q-q^{-1}) k^2 K^3}\ .\cr}\eqno(3.2)
$$
The measurable DCF is obtained by summing over both sectors, 
i.e.,:
$$
 S^{+-}(w,k-\pi)=S^{0,+-}(w,k-\pi)+S^{1,+-}(w,k-\pi)
 \ .\eqno(3.3)
$$
Finally, for fixed $w$ and $k$, $(\alpha_1,\alpha_2)$ is a 
solution to the
energy-momentum conservation laws:
$$
 \eqalign{w&=e(\alpha_1)+e(\alpha_2)\ ;\cr
 k&=-p_1-p_2\ .\cr}\eqno(3.4)
$$

\bigskip

\line{\bf\quad 3.a The isotropic limit\hfil}
\bigskip
As it is clear from the two equations (3.1) and (3.2), 
the 2-spinon
contribution in the anisotropic model is still quite 
untractable for
practical purposes. However, certain physical limits of that 
result are
quite simple, in comparison. Here we briefly rederive the 
known result
of the isotropic limit [28]. Let $p=-q$ be the nome of the 
various theta
functions that appear in $S_2^{i,+-}$. The isotropic 
($XXX$) limit is 
obtained by first making the redefinitions:
$$
 \eqalign{\xi&=ie^{{\epsilon\beta\over i\pi}} ,\cr
 p&=e^{-\epsilon}\ ,\cr}\eqno(3.6)
$$
and letting $\epsilon\ra 0^+$. In this isotropic limit, the real 
number $\beta$ is the
appropriate spectral parameter. Also, in this limit, we
have the following asymptotic behaviors: (see also Ref. [27]
\footnote{*}{\eightrm There is misprint there which is 
corrected here.})
$$
 \eqalign{(p^x;p)_{\infty}&\sim {(p;p)_{\infty}(1-p)^{1-x}\over
 \Gamma(x)}\ ;\cr
 \theta_p(-p^x)&\sim (2 \pi)^{1/2}(1-p)^{-1/2}\ ;\cr
 \gamma_{\pm}(-\xi^2)&\sim {(p^4;p^4)_{\infty}(1-p^4)^{1/2\mp
 1/4}A_\pm(\beta)\over \Gamma(1/2\pm 1/4)A_\pm(i\pi/2)}\ ;\cr
 (-p\xi^2;p^2)_{\infty} (-p\xi^{-2};p^2)_{\infty}&\sim
 {(1-p^2)(p^2;p^2)^2_\infty \over \Gamma(1/2-\beta/\pi i)
 \Gamma(1/2+\beta/\pi i)}\ ,\cr}\eqno(3.7)
$$
where:
$$
\eqalign{
 A_{\pm}(z)&=\exp\left( -\int_{0}^{\infty}dt {
 \sinh^2 t(1-{z \over i\pi})
 \exp(\mp t)\over t \sinh(2 t)\cosh(t)}\right),\cr 
|A_{\pm}(z)|^2&=
 \exp\left( -\int_{0}^{\infty}dt {(\cosh(2 t(1-{y\over\pi}))
 \cos({2 t x \over \pi})-1)
 \exp(\mp t)\over t \sinh(2 t)\cosh(t)}\right)\ ,\cr}\eqno(3.8)
$$
with $z=x+iy$ and $x$, $y$ being real.
Using the fact the for generic $k$, the two pairs $(p_1,p_2)$
and $(p_2,p_1)$
satisfying the energy-momentum conservation relations
have equal contributions to $S_2^{i,+-}$, then this latter
reduces to the following expression [28]:
$$
 S^{+-}_2(w, k-\pi)=
 {1\over 4}{ 
 |A_-(\beta({p_1})-\beta({p_2})|^2\over
 \sqrt{w_u^2-w^2}}\ ,\eqno(3.9) 
$$
with:
$$
 \eqalign{
 w&=-\pi(\sin({p_1})+
 \sin({p_2}))\ ;\cr
 k&=-{p_1}-{p_2}\ ;\cr
 w_u&=2 \pi \sin(k/2)\ ;\cr
 w_l&=\pi |\sin k|\ .\cr}\eqno(3.10)
$$
Let us mention that $S_2^{+-}(w,k-\pi)$ has been expressed in 
[29] in terms of $w$ and $k$ through 
the following relation: 
$$
 \beta({p_1})-\beta({p_2})=
 2\cosh^{-1}\sqrt{{w_u^2-w_l^2\over w^2-w_l^2}}\ .\eqno(3.11)
$$
Also, recall that $S_2^{i,+-}$ vanishes outside the domain 
limited by $w_u$
and $w_\ell$. Summing up over the two sectors and over the two
possible contributing pairs of spinons,  and using the relation:
$$
 \sigma^{\pm}={\sigma^x\pm i\sigma^y\over 2}\ ,\eqno(3.12)
$$ 
we find:
$$
 S^{\mu\mu}_2(w,k-\pi)=8S^{i,+-}_2(w,k-\pi),\quad \mu=x,y,z\ .
 \eqno(3.13)
$$
Note here that we have included the longitudinal component of 
the 
two-spinon DCF with the two transverse ones because they are 
all equal
in this isotropic limit. For further applications of this 
result, see [29].
In section 4, we discuss in detail the four-spinon DCF in
the isotropic limit.

\bigskip

\line{\bf\quad 3.b The Ising limit\hfil}
\bigskip
The other practically interesting limit of the anisotropic
DCF is the Ising limit, which is already known in the 
literature [33].
Here we briefly rederive it in the context of 
the present theory,
see also [30], and compare it to the perturbative result of 
Ishimura
and Shiba [33].

The Ising limit is the limit in which the interaction term
${1\over 2}\sum_n \sigma^z_n\sigma^z_{n+1}$ in the Hamiltonian
becomes dominant. It is therefore obtained by first rescaling the
Hamiltonian as $H_{\rm Ising}=H_{XXZ}/|\Delta|$, and then 
letting the nome
$-q=\epsilon\ra 0^+$. This means the energy $e$ of a spinon and
hence the factor $J(p_1,p_2)$ in eq (3.1) above must also be 
rescaled
as $e_{\rm Ising}=e/|\Delta|$ and $J_{\rm Ising}(p_1,p_2)=
J(p_1,p_2)/|\Delta|$, respectively.
Using also the following first order expansions in the 
limit $\epsilon \ra 0$:
$$
 \eqalign{K&\sim {\pi\over 2}(1+4\epsilon+O(\epsilon^2))\ ;\cr
 k&\sim 4\epsilon^{1/2}(1-4\epsilon+O(\epsilon^2))\ ;\cr}
\eqno(3.14)
$$
we find, for $\mu=x,y$ 
$$
 \eqalign{S^{\mu,\mu}_{\rm Ising}(w,k-\pi)&=
 2S^{0,+-}(w,k-\pi)+2S^{1,+-}(w,k-\pi)\ ;\cr
 &={\sqrt{4|V|^2-(w-2)^2}\over  |V|^2}\Big(1-4\epsilon \cos(k)
 -{(w-2)\big(1+\cos^2(k)\big)\over 2\cos^2(k)}\Big)\ ;\cr}
\eqno(3.15) 
$$
with:
$$
 \eqalign{e(p)&\sim 1-4\epsilon \cos(2 p)\ ;\cr
 w&=e(p_1)+e(p_2)\sim 2-8\epsilon \cos(k)\cos(k+2p_1)\ ;\cr
 V&=4\epsilon \cos(k)\ ;\cr
 |w-2|&\leq 2|V|\ . \cr}\eqno(3.16)
$$
Note that, except for the third one, all other factors in the 
above formula agree with those of Ishimura and Shiba [33].
These authors have used a completely different method,
i.e., the Green function technique. Also, note that both 
formulas completely
coincide in the particular cases of $w=2$ and $k=\pi$. 

The most important
difference is that in our expression, the order of the pole at
$k=\pi/2$ is increased by two at the zeroth order in $q$. 
It is our opinion
that the disagreement between the two formulas is 
due to the fact that the
authors of [33] have used a dispersion relation 
only up to first order
in $\ep$. However, following Ref. [21], 
the DCF can be written as:
$$
 S^{+-}(w,k-\pi)=D(w,k-\pi)\ M(w,k-\pi)/2\ ,\eqno(3.17)
$$
where we have:
$$
 \eqalign{D(w,k-\pi)&=2 |\Delta|/ |dw/dp_{1}|=
 {1\over 8|\sin(2p_1+k)\cos(k)\;\epsilon-2\sin(4p_1+2k)
 \cos(2k)\;\epsilon^2|}\cr
 &\sim\ {1\over 8\epsilon |\sin(2p_1+k)\cos(k)|}\
 (1+ 4{\cos(2p_1+k)\cos(2k)\over \cos(k)}\ \epsilon)\ .\cr}
\eqno(3.18)
$$
From the above relation, it is clear that a second order 
dispersion relation
in $\epsilon$ contributes to  zeroth order in the DCF. 
Therefore, it seems
that this second order term in the dispersion 
relation is missing in
the Ishimura-Shiba formula whereas it is 
present in our expression.

\bigskip\bigskip

\line{\bf 4. The four-spinon contribution to the DCF\hfil}

\bigskip
Let us now turn to the main subject of this paper, namely, 
the four-spinon contribution. As mentioned in the 
introduction, as far as
we know, no previous study of this case, whether 
numerical or analytic,
is available in the literature. 
After we process a little more the anisotropic case, 
we quickly focus our study on the isotropic case, and 
then we discuss the
Ising limit. As the reader will rapidly realize, the 
reason is that in
the anisotropic case, the expression of the four-spinon 
contribution,
while simplifying somewhat, is still difficult to manipulate.

We let $n=4$ in (2.3) and in the subsequent relations 
of section 2.
The set $\L$ has one element and so, there is one 
contour integral
to perform. The four-spinon contribution (that we denote $S_4$)
is defined in terms of the matrix elements
$X^i_{\ell}(\xi_4,\dots,\xi_1)$, where $\ell=1,\dots,4$
corresponds to the position of the $+$ sign in the sequence
$\{\epsilon_1,\epsilon_2,\epsilon_3,\epsilon_4\}$ that 
enters in the
definition of the set $\L$, equation (2.8). This matrix element
can be written as:
$$
 X^i_{\ell}={\prod_{1\leq j<k\leq 4}\gamma(u_k/u_j)\over
 \prod_{j=1}^4(-qu_j;q^4)_{\infty}(-q^3 u_j;q^4)_{\infty}}
 \prod_{j=1}^4\xi_j^{j-(1+\mu_j)/2-i}\ I_\ell\ ,\eqno(4.1)
$$
where:
$$
 \eqalign{I_\ell &=\oint {dv_\ell \over 2 \pi i v_\ell}
 \prod_{\ell<j}{u_j-v_\ell\over u_j v_\ell}
 \prod_{j<\ell}{qu_j-q^{-1}v_\ell\over u_j v_\ell}\cr
 &\times \prod_{j=1}^4{1\over ({u_j\over v_\ell};q^4)_\infty 
 (q^{-2}{v_\ell\over u_j};q^4)_\infty}
 (-q^{-1}v_\ell;q^2)_\infty (-q^{3}v_\ell^{-1};q^2)_\infty 
 G^{+,i}(v_\ell)\cr}. \eqno(4.2)
$$
The contour is as described in section 3, after 
equation (2.8) and 
$G^{+,i}(v_\ell)$ simplifies to:
$$
 \eqalign{
 G^{+,i}(v_\ell)&=(-v_\ell)^i \theta_{q^8}(-q^{-2+4i}
{v^2_\ell \over u_1 u_2
 u_3 u_4})C^+_{4,i},\cr
 C^+_{4,i}&=(-q)^{1-2i}(1-q^{-2})(q^2;q^2)_\infty
 (q^4;q^4)^2_\infty \rho^4(q),\cr}
 \eqno(4.3)
$$
Because all the poles are simple, it is therefore 
possible to perform the contour integral present in the above
expression of $I_\ell$. Being careful about the zeroes that 
annihilate
some of the poles, we find the following series expansion for it:
$$
 \eqalign{
 I_\ell^i &=\sum_{j=1}^4\sum_{m=\Theta(j-\ell)}
 u_j q^{4m^2-1}\prod_{i<\ell} (qu_j^{-1}-q^{4m-1}u_i^{-1})
 \prod_{i>\ell} (u_j^{-1}-q^{4m}u_i^{-1})\cr
 &\times \prod_{i\neq j}{1\over ({u_i\over u_j};q^4)_\infty 
 (q^{-2+4m}{u_j\over u_i};q^4)_\infty \prod_{s=1}^m 
 (q^{4s}-{u_i\over u_j})}\cr
 &\times {(-q^{-1+4m}u_j;q^2)_\infty (-qu_j^{-1};q^2)_\infty
 G^{+,i}(q^{4m}u_j)\prod_{s=0}^{2m-1}(q^{4m-1-2s}+u_j^{-1}) 
 \over
 (q^{4};q^4)_{\infty}  (q^{-2+4m};q^4)_{\infty}
\prod_{s=1}^{m}(q^{4s}-1)}\ .\cr} \eqno(4.4)
$$ 

The recipe to derive a series expansion for $S_4$ is to 
substitute this
expression for $I_\ell$ back into the expression of 
$X^i_\ell$, and
then substitute the latter in the expression of $S_4$. 
The general term
we thus find is quite complicated and therefore we prefer 
to omit reporting
it in this article. We rather focus our attention on the 
isotropic limit
(and later on we discuss the Ising limit) where major 
simplifications occur.

\bigskip

\line{\bf 4.1 The isotropic limit\hfil}
\bigskip
The isotropic limit is performed by first redefining 
$u_i=-\xi_i^2=e^{2\epsilon \beta_i/i\pi}$
and $p=-q=e^{-\epsilon}$ and taking the limit $\epsilon\ra 0^+$. 
Using  the following relations, which give the 
leading terms in this limit
of the different expressions involved in this calculation: 
$$
 \eqalign{
  &\rho^2(q)=\gamma_+(q^{-2})\ra\gamma(u)=\gamma_-(u)\ ;\cr
 &u=-\xi^2=e^{2\epsilon \beta\over \pi i}\ ;\cr
 &u_1-u_2 q^{4m}\sim {2\epsilon\over i\pi}
 (\beta_1-\beta_2+2\pi i m)\ ;\cr
 &u_1-u_2 q^{-2+4m}\sim {2\epsilon\over i\pi}
 (\beta_1-\beta_2+\pi i (2m-1))\ ;\cr
 &(-qu;q^4)_\infty (-q^3 u^{-1};q^4)_\infty\sim
 {(q^4;q^4)^2_\infty (1-q^4)\over \Gamma({1\over 4}-
 {\beta\over 2 \pi i})\Gamma({3\over 4}+{\beta\over 2\pi i})}\ ;
\cr
 &(-q^{-1+4m}u;q^2)_\infty (-q^{3-4m}u^{-1};q^2)_\infty
 \sim -(q^2;q^2)^2_\infty (1-q^2)\cosh(\beta)\ ;\cr
 &(1-q^{-2+4m})(q^{-4m};q^4)_{(m),\infty} 
 (q^{-2+4m};q^4)_{(m),\infty}\sim
 (-1)^{m+1}{m!(q^4;q^2)^2_\infty (1-q^4)^{1\over 2}\over
 \Gamma(m-{1\over 2})}\ ,\cr}\eqno(4.5)
$$
we find after a lengthy algerba that the isotropic ($XXX$) limit
of $S_4$ simplifies to:
$$
 S_4(w, k-\pi)= C_4 \int_{-\pi}^{0}dp_3 
 \int_{-\pi}^0 dp_4
 \sum_{(p_1,p_2)}{f(\beta_1,\beta_2,\beta_3,\beta_4) 
\sum_{\ell=1}^4 
 |g_{\ell}(\beta_1,\beta_2,\beta_3,\beta_4)|^2
 \over\sqrt{W_u^2 - W^2}}\ , \eqno(4.6) 
$$
a rather manageable expression. The notation is as follows:
$$
 \eqalign{&f(\beta_1,\beta_2,\beta_3,\beta_4)=
 \prod_{1\leq j<j^\prime\leq4}
 |A_{-}(\beta_{j^\prime}-\beta_j)|^2\ ;\cr
 &C_4={1  \over 3\times 2^9 \Gamma(1/4)^{8}
 |A_{-}(i\pi/2)|^{8}}\ ;\cr
 &W=w+\pi(\sin{p_3}+\sin{p_4})\ ;\cr
 &W_u=2 \pi |\sin(K/2)|\ ;\cr
 &K=k+p_3+p_4\ ;\cr
 &\cot{p_j}=\sinh{\beta_j}\ ,
\qquad -\pi\le p_j\le 0\  ,\cr} \eqno(4.7)
$$
and $g_\ell$ is to be defined shortly. 
Also, for fixed $W$ and $K$,
the sum $\sum_{(p_1,p_2)}$ is over the two pairs $(p_1,p_2)$ and
$(p_2,p_1)$ 
solutions
to the energy-momentum conservation laws:
$$
 \eqalign{W&=-\pi(\sin{p_1}+\sin{p_2})\ ;\cr
 K&=-p_1 -p_2\ ,\cr}  \eqno(4.8)
$$
and which are in fact given by
$$
(p_1, p_2) = 
 \Big(-{K\over 2}+\arccos\big({W\over{2\pi\sin{K\over2}}}
\big)\ ,\ -{K\over 2}-
 \arccos\big({W\over{2\pi\sin{K\over 2}}}\big)\Big)\ .
\eqno(4.9)
$$

The rest of this subsection is devoted to 
the discussion of some of the
properties of the integrand of $S_4$. 
We start by discussing the
function $g_\ell$, which is given by:
$$
 \eqalign{g_\ell = (-1)^{\ell+1} (2\pi)^4 \sum_{j=1}^4 
 \cosh(2\pi\rho_j)&
 \sum_{m=\Theta(j-\ell)}^{\infty} {{\prod_{i\ne\ell}
 (m-{1\over 2}\Theta(\ell-i)+i\rho_{ji})}\over
 {\prod_{i\ne j}\pi^{-1}\sinh(\pi\rho_{ji})}}\cr
 &\times\prod_{i=1}^4{{\Gamma(m-{1\over 2}+i\rho_{ji})}\over
 {\Gamma(m+1+i\rho_{ji})}}\ ,\cr}\eqno(4.10)
$$
where we have set $\beta_j/{2\pi}\equiv \rho_j$ and
$\rho_j-\rho_i\equiv\rho_{ji}$. It can be rewritten as:
$$
 \eqalign{g_\ell = (-1)^{\ell}(2\pi)^4 \sum_{j=1}^4 
{1\over\sin{p_j}}
 \prod_{i\ne j}\pi{\sqrt{\sin p_i \sin p_j}\over 
\sin\big({{p_j-p_i}
 \over 2}\big)}\ & \sum_{m=\Theta(j-l)}^{\infty}\
 \prod_{i\ne \ell} (m-{1\over 2}\Theta(\ell-i)+i\rho_{ji}) \cr &
 \times\ \prod_{i=1}^4 {{\Gamma(m-{1\over 2}+i\rho_{ji})}\over
 {\Gamma(m+1+i\rho_{ji})}}\ ,\cr} \eqno(4.11)
$$
expression in which we have used the following:
$$
 \eqalign{&\cosh(2\pi\rho_j) \sin p_j = -1\ ;\cr
 &\sinh(\pi\rho_{ji})=
 {{\sin\big({{p_j-p_i}\over 2}\big)}\over
 \sqrt{\sin p_i\sin p_j}}\ ;\cr
 &\cosh(\pi\rho_{ji}) =
 -{{\sin\big({{p_i+p_j}\over 2}\big)}\over
 \sqrt{\sin p_i\sin p_j}}\ .\cr} \eqno(4.12)
$$
Both expressions for $g_\ell$ are useful in the sequel. 

We first discuss the behavior of $g_\ell$ in the region
where two $\beta_i$'s (equivalently two $p_i$'s) are equal.
We find that $g_\ell$ is finite for all $\ell$.
The same is true in the regions where three of the $\beta_i$'s
or all the four are equal. Note that for a given 
$g_\ell$, each term in the $\sum_{j=1}^4$ can be 
divergent, but when  
the divergent pieces are put together, they cancel one another. 
To be specific, consider for example $g_1$ in the region 
$\rho_{12}$ very small
(the other $\rho_{ji}$ being kept finite). Then we have:
$$
 \eqalign{g_1/(2\pi)^4 &= {1\over\rho_{12}}\cosh(2\pi\rho_1)
 \sum_{m=1}^{\infty}m\pi^2{{(m+i\rho_{13})(m+i\rho_{14})}\over
 {\sinh(\pi\rho_{13})\sinh(\pi\rho_{14})}}
 \prod_{i=1}^4{{\Gamma(m-{1\over 2}+i\rho_{1i})}\over
 {\Gamma(m+1+i\rho_{1i})}}\cr
 &+{1\over\rho_{21}}\cosh(2\pi\rho_2)\sum_{m=1}^{\infty}m\pi^2
 {{(m+i\rho_{23})(m+i\rho_{24})}\over{\sinh(\pi\rho_{23})
 \sinh(\pi\rho_{24})}}\prod_{i=1}^4{{\Gamma(m-{1\over 2}
+i\rho_{2i})}
 \over{\Gamma(m+1+i\rho_{2i})}}\cr
 &+{\rm regular}\ .\cr}\eqno(4.13)
$$
Now replacing $\rho_2$ by $\rho_1$ in  any regular part and 
using $\rho_{12}=-\rho_{21}$, we find that the singular pieces 
cancel each other and the resulting nonvanishing function 
is regular. The same is true in all the other cases.\footnote{*}
{\eightrm Since $\ninemi\ninesy |A_{-}|^2$ 
goes to zero in these regions, 
see [29], then the whole integrand of 
$\ninemi\ninesy S_4$ has a nice
regular behavior there.}

We now turn our attention to the asymptotic 
behavior of $a_{\ell m}$,
the general term of the series in $g_\ell$. 
The behavior of $a_{\ell m}$
at large $m$ can be obtained if we use the  asymptotics of the
$\Gamma$-function. We find that at large $m$:
$$
 \eqalign{&\Gamma(m-{1\over 2}+i\rho_{ji})\  \sim\ 
\Gamma(m-{1\over 2})\ 
 \sim\ \sqrt{2\pi} \exp{[(m-1)\ln{(m-{1\over 2})} -m+ 
{1\over 2}]}\ ;\cr
 &\Gamma(m+1+i\rho_{ji})\ \sim\ \Gamma(m+1)\ \sim\ 
 \sqrt{2\pi} \exp{[(m+{1\over 2})\ln{(m+1)} -m-1]}\ ;\cr
 &\prod_{i\ne\ell} (m-{1\over 2}\Theta(\ell-i)+i\rho_{ji})\  
\sim\ m^3\ .\cr}
 \eqno(4.14)
$$
We gather all the pieces together and we arrive at the following
asymptotic behavior:
$$
 a_m \ \sim\ {{(-1)^{\ell+1}(2\pi)^4}\over m^3}\sum_{j=1}^4 
 {\cosh(2\pi\rho_j)\over\prod_{i\ne j} \pi^{-1}
\sinh(\pi\rho_{ji})}\ ,
 \eqno(4.15)
$$
which leads to $\lim_{m\to\infty}|a_{m+1}/ a_{m}|=1^{-}$.
The series is thus convergent for generic $\rho_i$, 
though  not `very' rapidly. 

Finally, we discuss the behavior of the 
integrand of $S_4$ when one of the
four $\rho_i$'s goes to $\pm\infty$ 
(equivalently, $p_i\to -\pi$ or 0).
To be specific, and because in $S_4$ 
there is an explicit integration
over $p_4$, we consider
explicitly the case $\rho_4\rightarrow +
\infty$ while keeping the
others finite. Note that the conclusion 
we arrive at is the same for all
the other momenta. We first discuss the 
behavior of $g_\ell$ and then
we incorporate $f$ into the picture and discuss the behavior
of $f\ \sum_{\ell =1}^4 |g_\ell|^2$.

To start, let us consider for example $g_1/(2\pi)^4$. 
Explicitly, it is equal to:
$$
 \eqalign{g_1/(2\pi)^4&=\cosh(2\pi\rho_1)\sum_{m=0}^{\infty}
 {{\prod_{i=2}^4(m+i\rho_{1i})}\over{\prod_{i\ne 1}\pi^{-1}
 \sinh(\pi\rho_{1i})}} \prod_{i=1}^4 {{\Gamma(m-{1\over 2}+i
 \rho_{1i})}\over
 {\Gamma(m+1+i\rho_{1i})}}\cr
 &+\cosh(2\pi\rho_2)\sum_{m=1}^{\infty}{{\prod_{i=2}^4(m+i
 \rho_{2i})}\over
 {\prod_{i\ne 2}\pi^{-1}\sinh(\pi\rho_{2i})}} \prod_{i=1}^4
 {{\Gamma(m-{1\over 2}+i\rho_{2i})}\over{\Gamma(m+1+i
 \rho_{2i})}}\cr
 &+\cosh(2\pi\rho_3)\sum_{m=1}^{\infty}{{\prod_{i=2}^4
 (m+i\rho_{3i})}\over
 {\prod_{i\ne 3}\pi^{-1}\sinh(\pi\rho_{3i})}} \prod_{i=1}^4
 {{\Gamma(m-{1\over 2}+i\rho_{3i})}\over{\Gamma(m+1+i
 \rho_{3i})}}\cr
 &+\cosh(2\pi\rho_4)\sum_{m=1}^{\infty}{{\prod_{i=2}^4
 (m+i\rho_{4i})}\over
 {\prod_{i\ne 4}\pi^{-1}\sinh(\pi\rho_{4i})}}\prod_{i=1}^4
 {{\Gamma(m-{1\over 2}+i\rho_{4i})}\over{\Gamma(m+1+i
 \rho_{4i})}}\ .\cr}
 \eqno(4.16)
$$
For illustration, denote the term 
containing $\cosh(2\pi\rho_1)$ by the
superscript (1). Using the asymptotic behavior of 
$\Gamma(z)$ at large $|z|$, we have:
$$
 {{\Gamma(m-{1\over 2}+i\rho_{14})}\over{\Gamma(m+1+i
 \rho_{14})}}\ 
 \sim\ {1\over{(m+1+i\rho_{14})^{3\over 2}}}\ .\eqno(4.17)
$$
We then re-write this first term (1) as:
$$
 {{\cosh(2\pi\rho_1)}\over{\pi^{-1}\sinh(\pi\rho_{14})}}
 \sum_{m=0}^{\infty}{{(m+i\rho_{12})(m+i\rho_{13})}\over
 {\pi^{-1}\sinh(\pi\rho_{12})\pi^{-1}\sinh(\pi\rho_{13})}} 
 \prod_{i=1}^3
 {{\Gamma(m-{1\over 2}+i\rho_{1i})}\over{\Gamma(m+1+i\rho_{1i})}}
 {{(m+i\rho_{14})}\over{(m+1+i\rho_{14})^{3\over 2}}}\ . 
\eqno(4.18)
$$
The series above still converges as $1/m^3$, thus keeping 
only a finite number of terms in it, 
we can neglect $m$ in front of 
$\rho_{14}$ and we get the asymptotic 
behavior of this first term:
$$
 g_1^{(1)} \ \sim\  \rho_4^{-{1\over 2}}\  e^{-\pi\rho_4}\ R \ .
 \eqno(4.19)
$$
(We have neglected $\rho_1$ in front of 
$\rho_4$ and we have extracted the
relevent exponential from the $\sinh$ term). 
Here the function $R$ is 
a generic notation for whatever is 
known to be regular. We work out
the other three terms in the 
expression of $g_1/(2\pi)^4$ in a similar
manner and we find that they behave 
asymptotically in the same way as this
first term behaves (except the fourth 
one which goes to zero faster).
Thus we can write:
$$
 g_1\ \sim\ \rho_4^{-{1\over 2}}\  e^{-\pi\rho_4}\ R\ .
\eqno(4.20)
$$
In fact, the same pattern happens for the other 
$g_\ell$'s and we 
arrive at:
$$
 \sum_{\ell =1}^4 |g_{\ell}|^2\ \sim\ {1\over\rho_4} 
e^{-2\pi\rho_4}\ R\ . 
 \eqno(4.21)
$$
Equivalently, the asymptotic behavior of 
$g_\ell$ can be derived in the
$p$-variables, and in the limit $p_4\to -\pi$, we find:
$$
 \sum_{\ell =1}^4 |g_\ell|^2\ \sim\ {-\sin p_4 \over 
\ln(-\sin p_4)}\ R\ .
 \eqno(4.22)
$$

Now we put into the picture the contribution from $f$. 
Using the notation of [29], $f$ is given by:
$$
 f = \exp[-I(t_{12})-I(t_{13})-I(t_{14})-I(t_{23})-
 I(t_{24})-I(t_{34})]\ ,
 \eqno(4.23)
$$
where:
$$
 -I(t)= C_2/2+\ln\big(t^{1/2} \sinh{{\pi t}\over 4}\big)+
 h(t)/2\ ,\eqno(4.24)
$$
and $t_{ji}\equiv 4\rho_{ji}$\footnote{*}{Note that in Ref. [29],
there are few misprints regarding the definition of 
$I(t)$ that are
corrected here.}. One checks explicitly that:
$$
 h(t)\to 0\quad{\rm as}\quad t\to +\infty\ ,\eqno(4.25)
$$
which means that when $\rho_4\to +\infty$,
$$
 f\ \sim\ \rho_4^{3/2} e^{3\pi\rho_4}\ R\ ,\eqno(4.26)
$$
where we neglect the other $\rho$'s in front of $\rho_4$. Using
the behavior (4.23) above, we find:
$$
 f\ \sum_{\ell =1}^4 |g_{\ell}|^2\ \sim\ \rho_4^{1/2}\
 e^{\pi\rho_4}\ R\ . \eqno(4.27)
$$
Had we used the $p_i$ variables, we would have 
obtained the behavior:
$$
 f\ \sum_{\ell =1}^4  |g_\ell|^2\ \sim\
 \sqrt{\ln(-\sin p_4)\over -\sin p_4}\ R\ . \eqno(4.28)
$$
The convergence of the integrand is best seen if we use
$d{p_4} = -2\pi d\rho_4/\cosh{(2\pi\rho_4)}$ to write: 
$$
 dp_4\ f\ \sum_{\ell =1}^4 |g_\ell|^2\ \sim\
 \rho_{4}^{1/2}\ e^{-\pi\rho_4}\ R\ d\rho_4\ . \eqno(4.29)
$$
It is interesting to note that this asymptotic behavior of the
integrand  of $S_4$ resembles that
of $S_2$ in the vicinity of the lower boundary 
$w=w_\ell$, and which also
corresponds to the limit where one of the 
two momenta goes to 0 or 
$-\pi$.

\bigskip

\line{\bf\quad 4.b The Ising limit \hfil}
\bigskip
We finish our study by considering the behavior of $S_4$
in the Ising limit. This limit is performed as in the 2-spinon
case, namely by first rescaling the energies of the four 
spinons by 
$|\Delta|^{-1}$ and thus the Jacobian in the expression of $S_4$,
and then letting $q\to 0^-$. We return to equation (4.4) 
above where a series
expansion of $I_\ell^i$ is given. Because of the factor 
$\prod_{i<\ell} (qu_j^{-1}-q^{4m-1}u_i^{-1})$, it is clear that 
the term with $\ell=4$ and $m=0$ is the leading one in 
powers of $q$. 
Explicitly, we find:
$$
 \eqalign{
 I^{0}&\sim -q\sum_{j=1}^4{1\over u_1u_2u_3u_j}\prod_{i\neq j}
 {1\over (1-{u_i\over u_j})}+O(q^2)\ ;\cr
 I^{1}&\sim q {u_4\over u_j^2}\prod_{i\neq j}
 {1\over (1-{u_i\over u_j})}+O(q^2)\ .\cr}\eqno(4.30)
$$ 
Furthermore, from the definition of $\alpha$ in terms of $\xi$,
see equation (1.5), we have the following relation 
between the physical
momenta $p_i$ and the spectral parameters 
$\alpha_i$ (up to first order in
$q$): 
$$
 \alpha_i= p_i-2q\sin(2 p_i)-{\pi\over 2}\ . \eqno(4.31)
$$ 
Now, using the following expansions:
$$
 \eqalign{
 \gamma_-({\xi_1^2\over \xi_2^2})  
 \gamma_-({\xi_2^2\over \xi_1^2})&\ \sim\  
4\sin^2(\alpha_1-\alpha_2)+O(q^2) ,\cr
 {(q^2;q^2)^4_\infty\over \theta_{q^2}(q\xi_1^2)
 \theta_{q^2}(q\xi_2^2)}&\ \sim\  1 +O(q) ,\cr}\eqno(4.32)
$$
We find that  $h^0+h^1$, which appears in the numerator of the 
integrand of $S_4$, greatly simplifies to: 
$$
 \eqalign{
 h^0+h^1=2^8 q^2\sum_{m,n=1}^4 (-1)^{m+n-2}&
 {\cos(p_m-p_n)\sin(p_{\bar m}-p_{\bar n})\sin(p_m+k/2)
\sin(p_n+k/2)
 \over
 \sin(p_{\tilde m}-p_{\tilde n})}\cr
 &\times \prod_{1\leq j< j^\prime \leq 4}
\sin(p_j-p_{j^\prime})\ ,\cr}
 \eqno(4.33)
$$ 
where:
$$
 \eqalign{
 &\tilde m=min(m,n)\ ;\cr
 &\tilde n=max(m,n)\ ;\cr
 &\bar m= min(\{1,2,3,4\}/\{m,n\})\ ;\cr
 &\bar n= max(\{1,2,3,4\}/\{m,n\})\ .\cr}\eqno(4.34)
$$
For example if $(m,n)=(1,3)$ then 
$\tilde 1=1$, $\tilde 3=3$, $\bar m= 2$, and $\bar n= 4$.
Moreover, using the following 
expansion of the denominator of the integrand of $S_4$: 
$$
 \eqalign{&{C\over \sin(2p_1)\sqrt{1-k^2\cos^2(p_2)}-
 \sin(2p_2)\sqrt{1-k^2\cos^2(p_1)}}\ \sim\ \cr
 &\hs60pt -{3\over 2q\cos(p_1+p_2)\sin(p_1-p_2)}
 (1-4q{\cos(p_1-p_2)\over \cos(p_1+p_2)})\ ,\cr}\eqno(4.35)
$$
we arrive finally at the following simple 
integral for the leading term in $q$ of
$S_4$:
$$
 \eqalign{
 S_4\sim {96 q \over \pi^2} \sum_{m,n=1}^4 (-1)^{m+n-2}
 \int_{-\pi}^0 \int_{-\pi}^0 dp_3 dp_4 & 
 {\sin(p_{\bar m}-p_{\bar n})\sin(p_m+k/2)\sin(p_n+k/2)
 \over
 \sin(p_{\tilde m}-p_{\tilde n})|\sin(p_1-p_2)\cos(p_1+p_2)|}\cr
 &\times \cos(p_m-p_n)
\prod_{1\leq j< j^\prime \leq 4}\sin(p_j-p_{j^\prime})
 \ .\cr}\eqno(4.36)
$$
Let us recall that $p_1$ and $p_2$ are given in terms of 
$p_3$ and
$p_4$ through the energy-momentum conservation laws.
Consequently, the ratio of the contribution of 
four spinons to that of
two spinons is of the order of $q^2$, thus 
confirming all previous numerical
studies which show the dominance of the 
contribution of two spinons.    

\bigskip\bigskip

\line{\bf 5. Conclusion\hfil}
\bigskip
In this paper, we have studied the exact spinon 
contributions to the
dynamical correlation function (DCF) of the anisotropic $XXZ$
Heisenberg model in the antiferromagnetic regime,
and their isotropic and Ising limits.

First we have given a general contour-integral representation for
the $n$-spinon contribution, which we have used 
to rederive most of the presently known
exact results on the two-spinon contribution $S_2$
in the $XXZ$ case and its $XXX$ and Ising limits.

Then we have focused our attention on the first nontrivial case
for which no results (even numerical ones related to finite-chain
calculations) are available yet, namely, the four-spinon 
contribution
$S_4$. In this context, we have shown that the contour integrals
can be performed, yielding mathematically well defined 
expressions,
free from divergences. This important result demonstrates
that the usual criticism against the methods based on
the quantum affine symmetry is unjustified. In fact, our results
can be thought of as yet another instance of the usefulness
of the quantum affine symmetry of the $XXZ$ model.

Finally, we touched upon the Ising limit of $S_4$.
We have shown that its leading contribution is
of the order of $q$ and thus, in this limit, the ratio of the 
four-spinon
contribution to that of the two-spinon one is of the order of 
$q^2$.
It seems that the expression of $S_4$, up to this order in 
the Ising limit,
is the simplest result possible for the DCF beyond two-spinon,
and thus we think it deserves further attention because
it might provide new insights on the behavior of the highly 
nontrivial
quantum mechanical nature of the Heisenberg model that is 
partly encoded
in $S_4$. Moreover, in this isotropic limit, we have found 
that though
the series defining $S_4$ converges, it does so quite
slowly. Therefore, we think further numerical analysis
is required to see what is  the best criterion of 
truncation so that
we get a satisfactory numerical evaluation, in view of any 
potential
application to new experiments where the main focus is to be
on the explicit detection and measure of any signature from four 
spinons.
Indeed, as far as we know, no experiment thus far has
effectively focused on this contribution. 
The main reason is that up to now, much of the attention of the
theoretical work has been limited to the two-spinon sector, 
and even in this
latter, our theoretical understanding has not been complete until
we have started to exploit systematically the quantum affine 
symmetry
of $XXZ$ [23], thus allowing the derivation of the simple 
exact result
of [28].

From a general perspective, it would certainly 
be interesting to extend
these methods to other models known to 
be invariant under quantum affine
symmetry, or any other 
infinite-dimensional symmetry. This is already
extensively implemented in conformal field theories. 

\bigskip

\line {\bf Acknowlegments\hfil}
The work of A.H.B. is supported by the NSF Grant \# PHY9309888.

\vfil\eject

\line{\bf References\hfil}

\bigskip

\item{1.} W. Heisenberg, {Z. Phys.} {\bf 49}, 619 (1928).

\item{2.} H. Bethe, {Z. Phys.} {\bf 71}, 205 (1931).

\item{3.} L. Hulth\'en, {Arkiv Mat. Astron. Fysik A11} 
{\bf 26}, 1 (1938).

\item{4.} E.H. Lieb and D.C. Mattis, {J. Math. Phys.} 
{\bf 3}, 749 (1962).

\item{5.} J. des Cloizeaux and J.J. Pearson, {Phys. Rev.}
          {\bf 128}, 2131 (1962).

\item{6.} R.B. Griffiths, {Phys. Rev. } {\bf 133}, A768 (1964). 

\item{7.} C.N. Yang and C.P. Yang, {Phys. Rev. } 
{\bf 150}, 321 (1966);
{\bf 150}, 327 (1966); {\bf 151}, 258 (1966).

\item{8.} Th. Niemeijer, { Physica} {\bf 36}, 377 (1967).

\item{9.} E. Barouch, B.M. McCoy and D.B. Abraham, 
{Phys. Rev.} {\bf A4}, 2331 (1971).

\item{10.} M. Gaudin, {Phys. Rev. Lett.} {\bf 26}, 1301 (1971).

\item{11.} M. Takahashi, {Prog. Theor. Phys.} {\bf 46}, 
401 (1971).

\item{12.} L.A. Takhtajan and L.D. Faddeev, {Russ. Math. Surveys}
           {\bf 34}, 11 (1979).

\item{13.} B.M. McCoy, J.H.H. Perk and R.E. Shrock {Nucl. Phys.}
           {\bf B220}, 35 (1983).

\item{14.} O. Babelon, H.J. de Vega and C.M. Viallet, 
{Nuc. Phys.}{\bf B220}, 13 (1983).

\item{15.} G. M\"uller and R. E. Shrock,
           { Phys. Rev.} {\bf B29}, 288 (1984).

\item{16.} J.M.R. Roldan, B.M. McCoy and J.H.H. Perk, { Physica }
           {\bf 136A}, 255 (1986).

\item{17.} V.E. Korepin, A.G. Izergin, and N.M. 
Bogoliubov,{\it The Quantum
           Inverse Scattering Method and Correlation Functions},
           Cambridge University Press, (1993).

\item{18.} F.H.L. Essler, H. Frahm, A.G. Izergin and V.E. 
Korepin,
  {\it Commun. Math. Phys.} {\bf 174}, 191 (1994).

\item{19.} V.E. Korepin, A.G. Izergin, F.H.L. Essler and D. 
Uglov,
   {\it Phys. Lett.} {\bf 190A}, 182 (1994).

\item{20.} A. Luther and I. Peschel, {Phys. Rev.} 
{\bf B12}, 2131 (1969).

\item{21.} G. M\"uller, H. Thomas, H. Beck, and J.C. 
Bonner, { Phys. Rev.}
          {\bf B24}, 1429 (1981).

\item{22.} A. Fledderjohann, M. Karbach, K.-H. M\"utter, and P. 
Wielath, {\it J. Phys.: Cond. Matter} {\bf 7} 8993 (1995).

\item{23.} O. Davies, O. Foda, M. Jimbo, T. Miwa, and A. 
Nakayashiki,
           {\it Comm. Math. Phys.} {\bf 151}, 89 (1993).

\item{24.} I.B. Frenkel and N.H. Jing, {\it Proc. Natl. Acad. 
Sci. }
           {\bf 85}, 9373 (1988).

\item{25.} A. Abada, A.H. Bougourzi and M.A. El Gradechi,
           {\it Mod. Phys. Lett.} {\bf A8}, 715 (1993).

\item{26.} A.H. Bougourzi, {\it Nuc. Phys.} {\bf B404}, 457
           (1993).

\item{27.} M. Jimbo and T. Miwa, {\it Algebraic Analysis of 
Solvable
           Lattice Models}, {American Mathematical Society, 
(1994)}.

\item{28.} A.H. Bougourzi, M. Couture and M. Kacir,
           {\it  Phys. Rev. } 
{\bf B54}, 12669 (1996).

\item{29.} M. Karbach, G. M\"uller and A.H. Bougourzi, 
 {\it Phys. Rev. B}, in Press.

\item{30.} A.H. Bougourzi, M. Karbach and G. M\"uller,
           {\it Two-spinon dynamic structure factor 
of the one-dimensional
           $S=1/2$ Heisenberg-Ising antiferromagnet}
           {\it Preprint ITP-SB-58-96}, 1996.

\item{31.} R.A. Weston and A.H. Bougourzi, 
{\it The dynamic correlation
           function of the XXZ model}, 
{\it Preprint CRM-2198}, 1994.

\item{32.}A.H. Bougourzi, {\it Mod. Phys. Lett.} {\bf B10},
1237 (1996). 

\item{33.} N. Ishimura and H. Shiba, {\it Prog. Theo. Phys.}
           {\bf 63}, 743 (1980).
\vfil\eject\end